\documentstyle[12pt]{article}

\begin{document}

\title{ Dark Matter and Cosmic Strings in Particle Models{\footnote{
Invited Plenary Talk Given at COSMO'99, ICTP, Trieste, Sept.27-Oct.2, 1999}}
}
\author{Xinmin Zhang\\
Institute of High Energy Physics, Academia Sinica\\
       Beijing 100039, P.R. China}



\maketitle

\begin{abstract}
 In this talk, I will discuss the mechanism of nonthermal 
production of the neutralino cold dark
matter from the decay of cosmic strings
 and the embedded defects, the $\pi$ and $\eta^\prime$ strings in the 
strong interaction sector of the standard model. 
\end{abstract}

\section{ Non-thermal Production of Neutralino Cold Dark Matter from
Cosmic String Decays }

To begin with, we consider a general case and calculate the relic mass 
density
of the lightest supersymmetric particle(LSP)\cite{Rachel}, then we will 
move on to a 
discussion of some implications. Consider a phase transition which is
induced by the vacuum expectation value (vev) of some Higgs field $\Phi$,
$\langle |\Phi |
\rangle = \eta$, and takes place at a
temperature $T_c$ with $T_c \simeq \eta$. The strings are formed by the Higgs
field $\Phi$
and gauge field $A$. 
 The
mass per unit length of the strings is given by $\mu = \eta^2$.
During the phase transition, a network of strings forms, consisting of both
infinite strings and cosmic string loops. After the transition, the 
infinite 
string network coarsens and more loops form from the
intercommuting of infinite strings. Cosmic string loops loose their 
energy by
emitting gravitational
radiation. When the radius of a loop becomes of the order of the string 
width, the loop releases its final energy into a burst of $\Phi$ and
$A$ particles. Those particles subsequently decay into LSP, which we 
denote by
$\chi$, with
branching ratios $\epsilon$ and $\epsilon'$. For simplicity we now
assume that all the final string energy goes into $\Phi$ particles. A single
decaying cosmic string loop thus releases $N \simeq 2 \pi \lambda^{-1}
\epsilon$ LSPs which we take to have a monochromatic distribution with energy
$E \sim {T^c \over 2}$ and $\lambda$ is the Higgs self coupling constant.

In such scenarios, we thus have two sources of cold
dark matter(CDM) which will contribute to the matter density of the universe.
We have CDM which comes from the standard
scenario of thermal production; it gives a contribution to the
matter density $\Omega_{therm}$. And we also have non-thermal
production of CDM which comes from the decay of cosmic string loops and gives
a contribution
$\Omega_{nonth}$. The total CDM density is $\Omega_{CDM} =
\Omega_{therm} + \Omega_{nonth}$. During the temperature interval between
$T_c$ and the LSP freezeout temperature $T_\chi$, LSPs
released by decaying comic string loops will thermalise very quickly
with the surrounding plasma, and hence will contribute to
$\Omega_{therm}$, which should not sensitively deviate from the value
calculated by the standard method.  However, below the 
LSP freezeout
temperature, since the annihilation of the LSP is by definition
negligible, each CDM particle released by cosmic string decays will
contribute to $\Omega_{nonth}$. We obviously must have
\begin{equation}
\Omega_{nonth}<1 \label{eq:bound}.
\end{equation}
This will lead us to a constraint (a lower bound) on the cosmic string
forming scale. We now calculate $\Omega_{nonth}$.

We assume that the strings evolve in the friction dominated regime so
that the very small scale structure on the strings has not formed  
yet. The network of strings can then be described by a single length
scale $\xi(t)$. In the
friction dominated period, the length scale $\xi(t)$ has been shown to
scale as \cite{RobRiotto}:
\begin{equation}
\xi(t) = \xi(t_c) \left ({t \over t_c}\right )^{3\over 2} \label{eq:xi}
\end{equation}
where $\xi(t_c) \sim (\lambda \eta )^{-1}$.
 The number density of cosmic string loops
created per unit of time is
   given by \cite{ShelVil}:
\begin{equation}
{dn\over dt} = \nu \xi^{-4} {d\xi \over dt}
\end{equation}
where $\nu$ is a constant of order 1. We are interested in loops
   decaying below $T_\chi$.
The number density of LSP released from $t_{lsp}$ till today is given by:
\begin{equation}
n^{nonth}_{lsp}(t_0) = N \nu \int^{\xi_0}_{\xi_F} \left ( {t \over t_0} 
\right
)^{3\over 2} \xi^{-4} d\xi \label{eq:no}
\end{equation}  
where the subscript $0$ refers to parameters which are
evaluated today.
$\xi_F = \xi(t_F)$ where $t_F$ is the time at which cosmic string
loops which are decaying at time $t_{\chi}$ (associated with the LSP 
freezeout
temperature $T_{\chi}$) have formed. Now the loop's average radius
shrinks at a rate \cite{ShelVil} ${dR\over dt} = - \Gamma_{loops} G \mu$,
where $\Gamma_{loops}$ is a numerical factor $\sim 10-20$. Since loops
form at time $t_F$ with an average radius $R(t_F) \simeq \lambda^{-1} G 
\mu   
M_{pl}^{1\over 2} t_F^{3\over 2}$, they have shrunk to a point at the time
$t \simeq \lambda^{-1} \Gamma^{-1}_{loops} M_{pl}^{1\over 2} t_F^{3\over 2}$.
Thus
$t_F \sim (\lambda\Gamma)^{2\over 3}_{loops} M_{pl}^{-{1\over 3}}
t_\chi^{2\over 3}$.
Now the entropy density is $ s = {2 \pi^2
\over 45} g_* T^3$ where $g_*$ counts the number of massless degrees of
freedom in the corresponding phase. The time $t$ and temperature $T$ are 
related by $t = 0.3 g_*^{-{1\over 2}}(T) {M_{pl}\over T^2}$ where 
$M_{pl}$ is the Planck
mass. Thus using Eqs.(\ref{eq:xi}) and (\ref{eq:no}), we find
that the LSP number density today released by decaying cosmic string
loops is given by:
\begin{equation}
Y^{nonth}_{LSP} = {n^{nonth}_{lsp}\over s} = {{6.75} \over {\pi}} 
\epsilon \nu
\lambda^2 \Gamma_{loops}^{-2} g_{*_{T_c}}^{-9 \over 4}
g_{*_{T_\chi}}^{3 \over 4} \,
 M_{pl}^2\, {T_{\chi}^4 \over T_c^6} \, , \label{eq:Ynonth}
\end{equation}
where the subscript on $g^*$ refers to the time when $g^*$ is evaluated.

The LSP relic abundance is related to $Y_{\chi}$ by:
\begin{eqnarray}
\Omega_\chi\, h^2 & \approx & M_{\chi} Y_{\chi} s(t_0) \rho_c(t_0)^{-1} 
h^2 \nonumber \\
  & \approx & 2.82 \times 10^8\, Y^{tot}_\chi\,
(M_{\chi}/{\rm GeV}) \label{eq:Omega}
\end{eqnarray}
where $h$ is the Hubble parameter and $M_\chi$ is the LSP mass.
Now $Y^{tot}_{LSP} =
Y^{therm}_\chi+ Y^{nonth}_\chi$; hence by setting $h=0.70$, Eqs.
(\ref{eq:Omega}) and (\ref{eq:bound}) lead to the following constraint: 
\begin{equation}
5.75 \times 10^8\, Y^{nonth}_\chi\, (M_{\chi}/{\rm GeV}) < 1. \label{eq:Yb}
\end{equation}  
We thus see that Eqs. (\ref{eq:Ynonth}) and (\ref{eq:Yb}) lead to a lower 
bound on the cosmic string forming temperature $T_c$.

Our results have important implications for supersymmetric extensions of the
standard model with extra $U(1)$'s (or grand unified models with an
intermediate $SU(3)_c \times SU(2)_L \times U(1)_Y \times U(1)'$ gauge
symmetry). Most importantly, the requirement $\Omega_{nonth} < 1$ imposes 
a new
 constraint on supersymmetric model building and rules 
 out many models with a low scale of a new symmetry breaking which 
produces defects such as cosmic strings.

Consider, for example, the model with an
extra $U_{B-L}(1)$ gauge symmetry. In this model,
 the strings will release not only right-handed neutrinos
$N_i$, but also their superpartners $\tilde N_i$. The heavy neutrinos 
$N_i$ and
their scalar partners $\tilde N_i$ can decay into various final states
including the LSP. The superpotential relevant to the decays is
$$ W = H_1 \epsilon L y_l E^c + H_2 \epsilon L y_\nu N^c ,$$
where $H_1, H_2, L, E^c$ and
$N^c$ are the chiral superfields and $y_l, y_\nu$ are Yukawa couplings 
for the
lepton and neutrino Dirac masses, $m_l = y_l v_1, m_D = y_\nu v_2$, with
$v_{1,2}$ being the
vacuum expectation values of the Higgs fields. At tree level,
the decay rates of $N_i$ into s-lepton plus Higgsino and lepton plus 
Higgs are
the same and they are smaller than the rate of $\tilde N_i$ decaying into
s-lepton plus Higgs and Higssino plus lepton by a factor of 2. If the 
neutralino is higgsino-like,
the LSP arise directly from the decays of the $N_i$ and $\tilde N_i$. If the
neutralino
is bino- or photino-like, subsequent decays of s-lepton into binos or 
photinos
plus leptons will produce the LSP. For
reasonable values of the parameters, we estimate the branching ratio 
$\epsilon$
of the heavy particle decay into LSP to be between 0.1 and 0.5. From Eq.
(\ref{eq:Ynonth}) it follows that string decays can easily produce
the required amount of LSP. However
too many LSPs will be generated unless the $B-L$ breaking scale,
$\Lambda_{B-L}$ is higher than about
$10^8$ GeV . In turn, this
will set a lower limit on the neutrino masses generated by the see-saw
mechanism, $m_\nu \sim m_D^2/ \Lambda_{B-L}$. Inserting numbers and taking
$m_D \sim m_\tau \sim 1.8$ GeV, one obtains that $m_\nu \leq 30$ eV.

Our lower limit on the $B-L$ symmetry breaking scale in gauged $B-L$ models
and in general models with an extra $U(1)$ \cite{Langacker} pushes the 
mass of
the new  gauge boson far above the Fermi scale,
rendering it impossible to test the new physics signals from the extra
$Z^\prime$ in accelerators.

As one more implication of our results, we consider the limit on the 
lifetime of the Z-string\cite{Vach} in the MSSM. Since the Z-string is 
produced 
during the electroweak phase transition, the stable Z-string would 
produce too much LSP and overclose the universe. However if the Z-string 
decays before the temperature of the LSP freezeout, then the LSP produced 
from the decay of the Z-string would be thermalized immediately, which 
would result in a negligible $\Omega_{nonth}$.

\section{$\pi$ and $\eta^\prime$ strings in QCD}
We consider an idealization of QCD with two species of massless
quarks $u$ and $d$. The lagrangian of strong interaction physics is 
invariant   under $SU_L(2) \times SU_R(2)$ (we will come back to the 
discussion of the $U_A(1)$ at end of this section) 
chiral transformations \begin{equation}
\Psi_{L, R} \rightarrow {\rm exp}(-i \vec{\theta}_{L,R} \cdot \vec{\tau} )
         \Psi_{L, R},
\end{equation}
where $\Psi^T_{L,R} = (u, d)_{L,R}$. However this chiral symmetry does
not appear in the low energy particle spectrum since it is spontaneously 
broken due to the quark condensate.
Consequently, three Goldstone bosons, the pions, appear and the (constituent)
quarks become massive.
At low energy, the spontaneous breaking of chiral symmetry can be 
described by  an effective theory, the linear sigma model, 
which involves the massless pions ${\vec \pi}$ and a massive $\sigma$ 
particle.

As usual, we introduce the field
\begin{equation}
\Phi = \sigma \frac{\tau^0}{2} + i {\vec \pi} \frac{\vec \tau}{2},
\end{equation}
where $\tau^0$ is unity matrix and $\vec\tau$ are the Pauli matrices with
the normalization condition $Tr(\tau^a \tau^b) =2 \delta^{ab}$. Under
$SU_L(2) \times SU_R(2)$ chiral transformations, $\Phi$ transforms as
\begin{equation}
\Phi \rightarrow L^+ \Phi R.
\end{equation}
The renormalizable effective lagrangian of the linear sigma model is
given by
\begin{equation}
{\cal L} = {\cal L}_{\Phi} + {\cal L}_{q},
\end{equation}
where
\begin{equation} \label{eq5}
{\cal L}_{\Phi} = Tr[ {( \partial_\mu \Phi)}^+ \partial^\mu \Phi] - \lambda
{[ Tr( \Phi^+ \Phi) - \frac{f_\pi^2}{2} ]}^2,
\end{equation}
and
\begin{equation} \label{eq6}
{\cal L}_q = {\bar \Psi}_L i \gamma^\mu \partial_\mu \Psi_L + {\bar \Psi}_R
       i \gamma^\mu \partial_\mu \Psi_R -2 g {\bar \Psi}_L \Phi \Psi_R
+ {\it h.c.}. 
\end{equation}
During chiral symmetry breaking, the field $\sigma$ takes on a nonvanishing
vacuum expectation value, which breaks $SU_L(2) \times SU_R(2)$
down to $SU_{L+R}(2)$. This results in a massive sigma
$\sigma$  and three massless Goldstone bosons
${\vec \pi}$, as well as giving a mass $m_q = g f_\pi$ to the
constituent quarks. Numerically, $f_\pi \sim 94$ MeV
and $m_{q} \sim 300$ MeV.
  
 We studied the classical solutions of this model
and  discovered a class of vortex-like configurations which we refer to 
as 
pion string\cite{Zhang}. Like
the Z string\cite{Vach} of the standard electroweak model, 
 the pion string is not topologically stable. Nevertheless, as 
demonstrated 
recently in numerical simulations in the case of semilocal 
strings\cite{Borrill} (which are also not topologically 
stable) pion strings are expected to be produced during the 
QCD phase transition in the early Universe (and also in heavy-ion 
collisions).   The strings will subsequently decay.

The pion string is a static configuration of the lagrangian
${\cal L}_\Phi$ of Eq. (\ref{eq5}).
To construct these solutions,  we define new fields
\begin{eqnarray}
\phi&=& \frac{ \sigma + i \pi^0}{ \sqrt 2 },\\
\pi^\pm &=& \frac{ \pi^1 \pm i \pi^2 }{ \sqrt 2}.
\end{eqnarray}
The lagrangian ${\cal L}_\Phi$ now can be rewritten as
\begin{equation}
{\cal L} =
{(\partial_\mu \phi )}^* \partial^\mu \phi
+ \partial_\mu \pi^+ \partial^\mu \pi^- - \lambda {( \pi^+ \pi^-
+ \phi^* \phi - \frac{f_\pi^2}{2} )}^2.
\end{equation}  
For static configurations, the energy functional corresponding to the 
above lagrangian is given by 
\begin{equation} \label{energy}
E = \int d^3 x
\left [ {\vec{\bigtriangledown} \phi }^* \vec{\bigtriangledown}\phi
+ \vec{\bigtriangledown}\pi^+ \vec{\bigtriangledown}\pi^- + \lambda
{( \pi^+ \pi^- + \phi^* \phi - \frac{ f_\pi^2}{2} )}^2 \right ].
\end{equation}
The time independent equations of motion are:
\begin{eqnarray}
 {\bigtriangledown}^2 \phi &=&  2 \lambda ( \pi^+ \pi^-
+ \phi^* \phi - \frac{f_\pi^2}{2} ) \phi,\\
 \bigtriangledown^2 \pi^+&=& 2 \lambda ( \pi^+ \pi^- + \phi^* \phi
   - \frac{f_\pi^2}{2} ) \pi^+ .
\end{eqnarray}
The pion string solution extremising the energy functional of Eq. 
(\ref{energy}) is given by
\begin{eqnarray}
\phi &=& \frac{ f_\pi}{\sqrt 2} \rho (r) e^{i  \theta}, \\
\pi^\pm &=& 0,  
\end{eqnarray}
where the coordinates $r$ and
$\theta$ are polar coordinates in the $x-y$ plane (the string is assumed 
to 
 lie along the z axis), and
$\rho(r)$ satisfies the following boundary conditions
\begin{eqnarray}
r \rightarrow 0, & & \rho(r) \rightarrow 0 ; \\
r \rightarrow \infty, & & \rho(r) \rightarrow 1.
\end{eqnarray}
The pion-string is not topologically stable, since any field configuration
can be continuously deformed to the vacuum.

In the discussions above we have neglected the electromagnetic interaction. 
When turning on $U_{em}(1)$, the derivatives are replaced by the 
covariant derivatives. Recently Nagasawa and Brandenberger\cite{Naga} have 
pointed 
out that the interactions with a finite temperature plasma will lead to 
corrections to the effective potential and may stabilize the $\pi$ strings.

Since a pion string is made of $\sigma$ and
$\pi^0$ fields, it is neutral under the $U_{em}(1)$ symmetry. However,
the $\pi^0$ will couple to photons via the Wess Zumino type interaction.
\begin{equation} \label{low}
{\cal L }_{low}=\frac{f_\pi^2}{4}Tr(\partial_\mu \Sigma^+ \partial^\mu 
\Sigma)
 - \frac{1}{4}F_{\mu\nu}F^{\mu\nu}
   -\frac{N_c \alpha}{24 \pi} \frac{\pi^0}{f_\pi}
   \epsilon^{\mu\nu\alpha\beta}F_{\mu\nu}F_{\alpha\beta},
\end{equation}
where $N_c = 3$, $\Sigma = exp(i {\vec\tau}\cdot{\vec\pi}/ f_\pi )$, and 
 $\alpha$ is the electromagnetic fine structure constant.

 Recently Brandenberger and I\cite{Bran} propose a new mechanism for the 
generation of primordial  
magnetic fields. It is based on the realization that
anomalous global strings couple to electricity and magnetism via the 
anomalous interactions shown above and induce magnetic fields . The major
advantage of this mechanism is that the coherence scale of the magnetic
fields induced by these global vortex lines is set by the length scale
$\xi(t)$ of the strings (the typical curvature radius of the strings).

Before concluding this section, I will speculate about
 the existence of an unstable $\eta^\prime$ string\cite{Zhang}.
In QCD, in the limit of massless quarks, there is an additional
 $U_A(1)$ chiral symmetry. This chiral symmetry,   
when broken by the quark condensate, predicts the existence of a goldstone
 boson. There is no such a light meson, however. This is resolved by
the Adler-Bell-Jackiw $U_A(1)$ anomaly together with the properties of
non-trivial
vacuum structure of non-abelian gauge theory, QCD. The
$U_A(1)$ symmetry is badly broken by instanton effects at zero temperature.

As the density of matter and/or the temperature increases, it is expected
that the instanton effects will rapidly disappear,
 and one thus has an additional $U_A(1)$ symmetry
 (besides $SU_L(2) \times SU_R(2)$)
at the transition temperature of the QCD chiral symmetry.
 When the $U_A(1)$ symmetry is broken spontaneously
by the quark condensate, a topological string, the $\eta^\prime$-string, 
results.
Differing from the pion-string, the $\eta^\prime$-string is topologically 
stable at high temperatures, but will decay as the temperature decreases.
The $\eta^\prime$-string can form during the chiral phase
transition of QCD. In the setting of cosmology, it will
 exist during a specific epoch below the QCD
chiral symmetry breaking temperature during the evolution of the
universe. In the context of heavy-ion collisions,
 it will exist in the plasma created by the collision.
The strings then become unstable as the temperature decreases and when
 the instanton effects become substantial.

\section*{Acknowledgments}
This work was supported in part by the NSF of 
China.

\end{document}